# A new framework for prognostics in decentralized industries: Enhancing fairness, security, and transparency through Blockchain and Federated Learning

T.Q.D. Pham[1], K.D. Tran[2,*], Khanh T. P. Nguyen[3], X.V. Tran[4], L. Köehl[5], and K.P. Tran[5]

[1]*University of Li`ege, MSM Unit, All´ee de la D´ecouverte, 9 B52/3, B 4000 Li`ege, Belgium*

[2]*International Research Institute for Artificial Intelligence and Data Science, Dong A University, Danang, Vietnam*

[3]*Universtity of Technology of Tarbes Occitanie Pyr´en´ees (UTTOP), Production Engineering Laboratory, 65000 Tarbes, France.*

[4]*Institute of Southest Vietnamese Research, Thu Dau Mot University, Binh Duong, Vietnam*

[5]*Univ. Lille, ENSAIT, ULR 2461 - GEMTEX - G´enie et Mat´eriaux Textiles, F-59000 Lille, France*

[*]*Corresponding author: K.D. Tran, ductk@donga.edu.vn*

**Abstract**

As global industries transition towards Industry 5.0, predictive maintenance (PM) remains crucial for cost-effective operations, resilience, and minimizing downtime in increasingly smart manufacturing environments. In this chapter, we explore how the integration of Federated Learning (FL) and blockchain (BC) technologies enhances the prediction of machinery's Remaining Useful Life (RUL) within decentralized and human-centric industrial ecosystems. Traditional centralized data approaches raise concerns over privacy, security, and scalability, especially as Artificial Intelligent (AI)-driven smart manufacturing becomes more prevalent. This chapter leverages FL to enable localized model training across multiple sites while utilizing BC to ensure trust, transparency, and data integrity across the network. This BC-integrated FL framework optimizes RUL predictions, enhances data privacy and security, establishes transparency, and promotes collaboration in decentralized manufacturing. It addresses key challenges such as (i) maintaining privacy and security, (ii) ensuring transparency and fairness, and (iii) incentivizing participation in decentralized networks. Experimental validation using NASA's CMAPSS dataset demonstrates the model's effectiveness in real-world scenarios, and we extend our findings to the broader research community through open-source code on GitHub, inviting collaborative development to drive innovation in Industry 5.0.

# 1 Introduction

In the rapidly evolving landscape of global industries, enterprises across sectors like aerospace, automotive, and manufacturing are experiencing significant transformations driven by Industry 5.0 [1,2]. While Industry 4.0 focuses on automation, data exchange, and cyberphysical systems, Industry 5.0 emphasizes the collaboration between humans and intelligent machines, fostering a more personalized, human-centered approach to manufacturing. This paradigm shift prioritizes resilience, sustainability, and human-machine symbiosis, with Artificial Intelligence (AI) and advanced technologies playing central roles in creating smart manufacturing environments.

AI's integration into smart manufacturing brings enhanced efficiency and predictive capabilities, enabling real-time monitoring, process optimization, and more effective decisionmaking. Within this context, predictive maintenance (PM) [3–5] has emerged as a key application, particularly in industries managing an extensive array of machines, each operating under distinct and dynamic conditions. Predictive maintenance not only reduces unplanned downtime and enhances cost-effectiveness, but it also aligns with the core principles of Industry 5.0, which focuses on adaptability, efficiency, and human-centered innovations.

However, the adoption of AI-driven PM introduces a significant challenge: the centralized data models that traditionally power AI [3,4]. Centralized models require vast amounts of data to be transferred to a central server for training. This poses concerns related to high costs, data privacy, and the specific operational conditions unique to each facility or organization. These issues become more pronounced as industries grow increasingly interconnected, making the need for decentralized data management solutions critical. In summary, these issues can be summarised as:

- **Data Privacy and Security**: Centralized data systems require sensitive information to be transferred and stored in a single location, increasing vulnerability to breaches and misuse. Securing this data without relying on centralized repositories is crucial in a decentralized industrial ecosystem.

- **Cost and Efficiency**: Transferring vast amounts of data to centralized servers incurs significant costs in terms of bandwidth and storage, which may negate the efficiency gains promised by AI.

- **Diverse Operating Conditions**: Centralized models struggle to account for the unique operating environments of individual manufacturing sites. Each factory or machine may face different stressors and maintenance needs, which are not easily generalized in a centralized framework.

Federated Learning (FL) [6,7] addresses these challenges by offering a decentralized approach to model training. Instead of transferring all data to a central server, FL allows each entity (i.e., a factory) to train predictive models on its local data. The model parameters, not the data itself, are then shared with a central server, where they are aggregated to create a more generalized model. This approach ensures data privacy, reduces transfer costs, and reflects the diverse operating conditions across all participating entities. However, as Industry 5.0 brings further complexity to manufacturing systems, even FL presents challenges [8–10]. These include managing diverse data formats, minimizing communication overhead, ensuring privacy during parameter exchanges, and accounting for the varied data characteristics of each participant.

An efficient way to resolve the problem is to integrate blockchain (BC) technology [11,12] into the FL framework. This advantages both technologies' potential benefits and addresses the complex obstacles that arise [13,14]. Including BC's immutable and transparent system within FL adds an extra layer of security, ensuring the integrity and verifiability of model updates and parameter exchanges across the distributed network of manufacturing sites. This, in turn, mitigates concerns over potential data breaches and unauthorized access, gaining trust within the collaborative ecosystem [15]. BC's decentralized nature also matches with FL's philosophy, as each site's locally trained models can be securely aggregated without a centralized authority. This integration facilitates secure and efficient model aggregation and amplifies the decentralized essence of FL itself [16,17].

Currently, BC-based FL in RUL prediction is still in its early stages. Particularly, BCbased FL are mainly focused on IoT and healthcare fields [18,19], which are out of scope with our work. Furthermore, these studies tend to concentrate on either intricate statistical and federated learning principles or on relatively complex industrial models, which may not be readily understandable for individuals who are new to the field.

Based on these backgrounds, the combination of BC and FL is particularly promising for Industry 5.0, where human-machine collaboration is key, and personalized, context-aware systems need to be built on secure, transparent, and decentralized infrastructures. BCempowered FL frameworks foster human-centric AI by enabling secure, adaptive learning across diverse manufacturing settings while maintaining strict privacy and security standards. Despite the early stage of BC-based FL applications in PM and RUL prediction, their potential is immense. While initial research has mainly focused on IoT and healthcare, this chapter aims to bring these advancements to smart manufacturing, bridging a gap in the existing literature. This chapter presents a novel framework that leverages the combined potential of FL and BC technologies to enable efficient RUL predictions in decentralized industries. The primary goal is to facilitate secure, privacy-preserving, and cost-effective PM while ensuring full transparency and trust. In doing so, we address the key challenges posed by centralized data models, bringing the decentralized essence of Industry 5.0 to PM systems. Moreover, we are committed to transparency and collaboration; thus, all the code related to this research is open-source on GitHub, inviting engagement and contributions from the broader community.

The chapter is structured as follows: Section 2 discusses the RUL prediction, covering both traditional and FL approaches while addressing the significant challenges inherent in this domain. Section 3 introduces the BC-based FL framework for RUL prediction to provide insights into its background, framework, and technical intricacies. Finally, a comprehensive experimental case study of our proposed BC-based FL framework is shown in Section 4 before the conclusion section.

# 2 Related works on RUL prediction

In this section, we provide an overview of RUL prediction, encompassing both traditional methods and the emerging FL approach. We then pivot to thoroughly examining the significant challenges that underlie RUL prediction in this context. Examining these challenges paves the way for introducing our innovative solution, leveraging the synergistic potential of FL and BC technologies to address these pressing issues effectively.

## 2.1 Traditional RUL prediction approaches at a single site

Traditional RUL prediction approaches typically rely on various aspects such as physical characteristics, structural attributes, historical data, sensor measurements, and the observed behaviors of components or systems to estimate the remaining operational lifespan at a single site. These conventional methodologies have been instrumental in predicting RUL, primarily emphasizing localized data analysis and predictive modelling. Specifically, data-driven approaches are increasingly favoured for RUL predictions, eclipsing modelbased methods due to their independence from knowledge of the underlying engineering system [3]. In industries, the inherent complexity of real-world systems makes it challenging to accurately represent all systems using solvable mathematical or physics-based approaches [4]. In addition, the proliferation of sensors has provided abundant data on system conditions. Concurrently, the maturation and widespread adoption of machine learning (ML) techniques have popularized ML-based models for RUL predictions, including approaches such as random forests and various neural networks. These models capture the intricate relationships between system health and RUL without necessitating an indepth understanding of the degradation processes.

However, data-driven prognostics in decentralized industries face multiple challenges due to the sparse nature of the data and the specific requirements of the predictive tasks [20]. One primary concern is the decentralized nature of the condition monitoring data, which is typically collected from diverse sites across the globe [21]. These data sets often present high variance due to the differing operational environments, machine usage patterns, and maintenance practices at each site [22]. As a result, building a generic model that can accurately predict the RUL across all sites becomes a considerable challenge [5,23]. Addressing these challenges requires a novel methodology that can leverage condition monitoring data from multiple sites while preserving each site's data privacy and trust. FL offers a promising solution to these challenges. In the next section, we discuss the FL-based approach for RUL prediction in detail.

## 2.2 FL-based approaches for RUL predictions across multiple sites

This section discusses how the FL overcomes problems with centralized methods. As a leader in collaborative machine learning, FL allows us to train models on many devices without sharing personal data. We will look at the basics, benefits, and uses of FL in RUL prediction. We will also talk about how FL can make a big difference in predictive maintenance by solving issues with old methods.

In an FL algorithm, a specific ML model is trained, distributed among some edge devices, and then aggregated in a server. Let $d$ be the dimension of the model parameters; FL algorithm finds a model parameter set as $w \in \mathrm{R}^d$ in each worker that can be expressed as:

$$\min_{w}\{f(w)\}, \tag{1}$$

where

$$f(\boldsymbol{w}) = \sum_{i=1}^{K} \frac{n_i}{K} f_i(\boldsymbol{w}) \tag{2}$$

is the loss function to be minimized, $K$ is the number of worker, $f_i(w)$ is the worker $k$'s loss function, and $n_i$ is size of worker's dataset. Various techniques such as learning federated averaging (FedAvg), federated stochastic gradient descent (FedSGD), and federated proximal (FedPro) can be used to facilitate collaborative model training across decentralized devices [24]. Particularly, FedAvg is lauded for its simplicity and interpretability but employs a fixed global learning rate. In contrast, FedSGD offers hyperparameter flexibility but may encounter privacy challenges. Finally, FedPro provides convergence guarantees under specific conditions but involves complex hyperparameter fine-tuning. In consideration of the objectives of our study, the FedAvg algorithm is chosen due to its simplicity and alignment with our research goals, focusing on FL's application in real-time RUL prediction, as detailed in [24].

In the FL-based approach, multiple ML algorithms can be used for RUL prediction, i.e., minimize loss function $f_i(w)$ between prediction and actual values. Multiple methods, including linear regression [25], radical basic functions [26], deep neural networks (DNN) [27], and support vector machine (SVM) [28] algorithms, can serve as ML models in the FL process. The figure 1 illustrates how ML models for RUL prediction are incorporated in the FL architecture across various sites. Each site (Factory 1 to Factory K) trains the RUL prediction models locally to maintain data privacy and security. These models' parameters are then transmitted to the aggregation server, which employs the FedAvg algorithm to integrate the individual contributions into a cohesive global model. Subsequently, this refined global model is redistributed to each site. Note that this chapter underscores the benefits of combining Blockchain (BC) technology with FL to enhance security and trust in the distributed learning process rather than improving the RUL prediction performance of the ML model utilized.

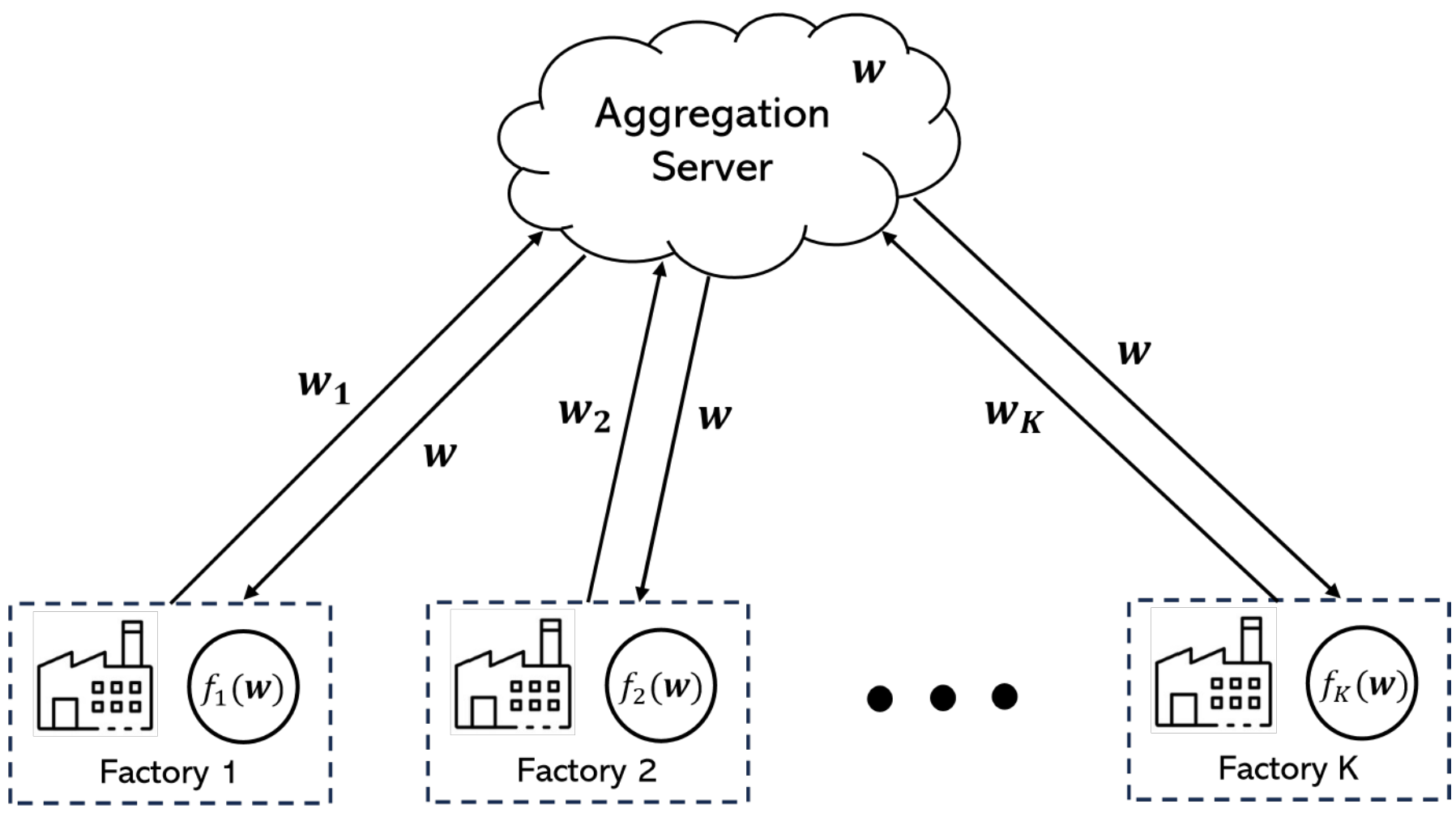

Fig. 1. Architecture of the FL process using ML model.

## 2.3 Major problems of FL in RUL prediction

In the FL domain, the quintessential challenge is to derive a comprehensive statistical model leveraging datasets dispersed across numerous remote devices, ranging from a few to potentially millions [8,9]. Such a process is encapsulated by Eq. (2), the local objective function $f_i(w)$ represents the empirical risk calculated from the data on each remote device. Although FL inherently integrates mechanisms designed to safeguard against unauthorized access and data breaches, it is not impervious to adversarial exploits, which could undermine the system's integrity and the data's confidentiality. As shown in Fig. 2, succinctly enumerates the advantages and disadvantages of employing FL in the context of real-time RUL predictions, delineating them into four beneficial attributes juxtaposed against four potential drawbacks.

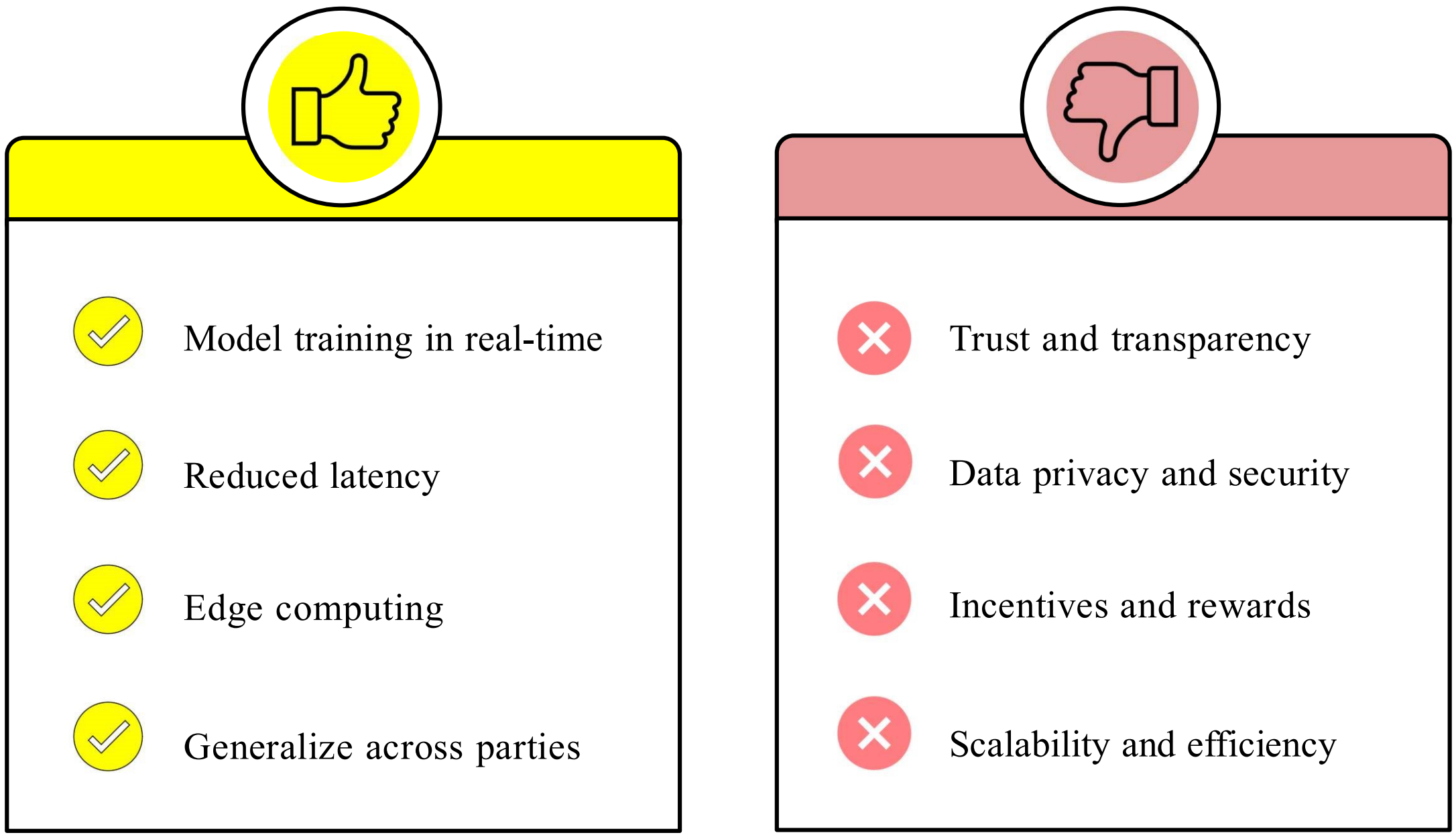

Fig. 2. Pros (left) and cons (right) of FL in RUL prediction.

Particularly, in the context of RUL predictions, it is essential to understand the four main drawbacks of FL and how they specifically impact this application.

1. **Trust and transparency:** They are fundamental in FL, especially for RUL predictions in critical industry sectors like aerospace and manufacturing. For collaborative model training across various factories or aircraft systems, it is essential to have a transparent process where each participant's contributions and model updates are verifiable. This ensures that the predictive models are reliable and their outputs are trusted for making crucial maintenance and operational decisions.

2. **Data privacy and security:** While FL enhances data privacy by keeping data on local devices, ensuring robust security in such a decentralized setup is challenging. In the context of RUL predictions, sensitive information about machinery health or aircraft system performance must be protected. Any breach could lead to industrial espionage or safety risks. Moreover, data integrity is vital to ensure accurate RUL predictions, as compromised data can lead to erroneous predictions, potentially causing safety hazards or operational inefficiencies.

3. **Incentives and rewards:** The lack of direct incentives in FL can impact the motivation of participants in manufacturing or aerospace industries to engage actively. These sectors require constant and accurate RUL predictions to maintain operational efficiency and safety standards. Participants might not contribute high-quality data or computational resources without proper incentives, leading to suboptimal RUL models.

4. **Scalability and efficiency:** In industries where numerous machines and components are constantly monitored for RUL predictions, the scalability of FL becomes a significant issue. The process of aggregating updates from a large number of devices can become resource-intensive and time-consuming. This can lead to delays in model updates or reduced model accuracy, adversely affecting the timeliness and precision of RUL predictions, which are crucial for maintenance scheduling and avoiding unexpected equipment failures.

In the pursuit of enhancing the efficacy of Federated Learning (FL) for industrial Remaining Useful Life (RUL) predictions, the integration of blockchain (BC) technology emerges as a promising solution. This amalgamation amplifies the inherent advantages of both FL and BC and adeptly navigates through the multifaceted challenges prevalent in this domain. The following section will delve into the intricacies of the proposed blockchainenhanced FL framework, tailored explicitly for RUL prediction in industrial contexts.

# 3 BC-based FL framework for RUL prediction

This section presents the BC-based FL framework designed for RUL prediction. It covers the fundamental components of BC, outlines the proposed framework, and delves into the technical details of how BC addresses key challenges within FL.

## 3.1 Block chain background

Indeed, BC technology presents a comprehensive remedy to the challenges above in FL. Establishing a decentralized and immutable ledger guarantees trust, transparency, and security throughout the collaborative model training process. Through mechanisms that reinforce data privacy, incentivize contributions, and bolster scalability, BC emerges as a pivotal enabler, ensuring FL's secure and efficient implementation in various dynamic and diverse environments. Hereafter, we discuss the essential ingredient of BC (i.e., definition and SC) and its architecture with FL.

**Blockchain:** It is essentially a decentralized distributed database [11]. All the interactive records (transactions) generated in the system are linked into chains as blocks and stored in each section in time. Furthermore, each transaction is guaranteed by cryptography and PoW algorithms that cannot be tampered with or forged, so each node in the system can achieve secure peer-to-peer transactions [13]. Fig. 3 illustrates the basic BC concept of BC and its data flow, including a block header containing metadata and some transaction records. These blocks are linked by the hash pointer of the block header to form a complete ledger, which is the narrow definition of BC. More precisely, from the bottom to the top, the BC comprises the data layer, incentive mechanism, consensus layer, network layer, and application layer.

The BC is generally divided into public, consortium, and private chains based on different application scenarios and designed systems. Generally, different types of BC are selected according to the requirements of different business scenarios. However, in a broad sense, only the public chain can meet the BC's original design intention.

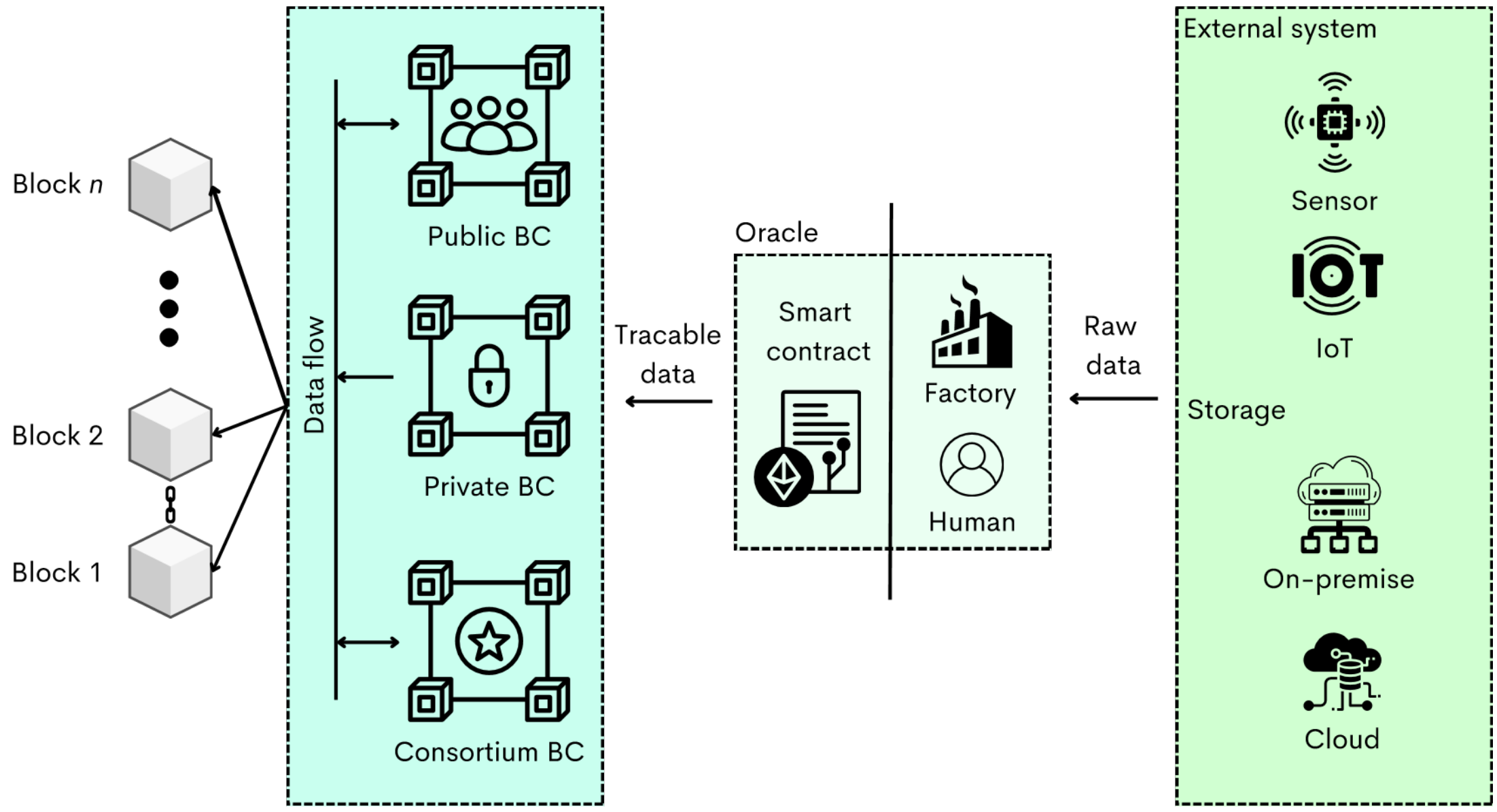

Fig. 3. Basic concept of BC and its data flow from raw to block.

**Smart contract (SC):** It is a critical application within BC. These contracts play a crucial role by digitally verifying negotiations or executions and facilitating trusted transactions without requiring third-party intervention. As depicted in Fig. 3, smart contracts include a mechanism for processing and preserving transactions. This versatility and adaptability extend to the BC environment, serving as the foundational infrastructure for deploying various algorithms, including FL. The digital verification of negotiated or executed contracts by SC ensures trustworthy transactions, eliminating the need for intermediaries. Notably, these transactions are characterized by traceability and irreversibility, enhancing the overall security and transparency of the process.

## 3.2 Proposed framework of BC-based FL for RUL predictions

BC is the foundational element in establishing a fully decentralized and secure architecture tailored for FL systems in our proposed framework. This integration offers resilient mechanisms for the protection of sensitive data, the cultivation of trust among stakeholders, and the mitigation of security vulnerabilities. Consequently, it finds particular relevance in applications associated with remote monitoring of machinery, detecting anomalies, diagnosing faults, and, notably, RUL) for vital and safety-critical systems.

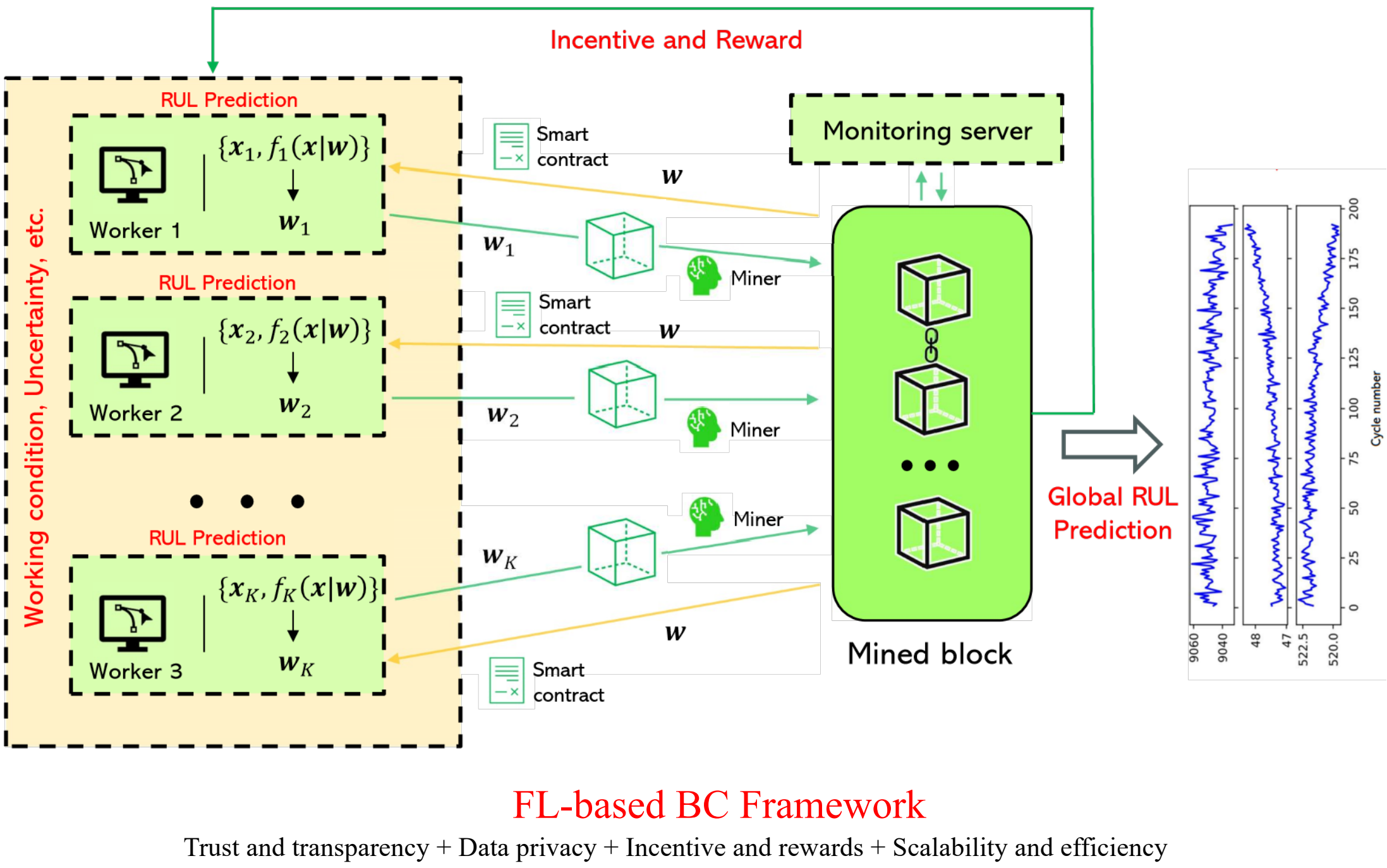

Fig. 4. Proposed architecture of BC-based FL for RUL prediction.

In the proposed BC-integrated FL framework for RUL predictions, depicted in Figure 4, the architecture is meticulously structured around six fundamental components: (1) worker *i*, (2) smart contract, (3) active miners $M_j$, (4) mined block, (5) consensus algorithm, and (6) the monitoring server. These components synergistically operate to optimize RUL predictions across the intricate landscapes of complex landscapes of various industry sectors, including manufacturing, automobile, aerospace, etc.

(1) Worker *i*: Each worker represents a participating entity in the FL process. It keeps the local data, like the condition monitoring records from a factory machine or an aircraft, and helps improve prediction models. This way, it ensures that the specific details of RUL in different working conditions are understood and considered.

(2) Smart Contract: It functions as the digital backbone of the FL framework, encoding the rules and procedures that govern the collaborative effort. In RUL predictions, the SC could, for example, automatically validate the accuracy of data from a jet engine's sensors. If the data meets predefined criteria, the model will be updated. It also manages the distribution of rewards, incentivizing factories or aerospace systems to provide high-quality data by granting them tokens or credits in return. The SC is the impartial enforcer that ensures all actions, from data submission to model updating, adhere to the established protocols.

(3) Active Miners $M_j$: Miners in the blockchain act as the gatekeepers of data integrity and security. In the context of RUL predictions, miners verify the sensor data or predictive model updates submitted by each worker before they are permanently recorded on the blockchain. This step is critical to prevent the introduction of false or malicious data that could lead to incorrect RUL predictions, which might result in premature component failure or unnecessary maintenance downtime.

(4) Mined Block: Each mined block is a collection of these verified updates, akin to a time-stamped batch of RUL insights from across the network, which could be parameters of prognostic models. The unique cryptographic hash within each block acts as a seal, cementing its place in the blockchain's ledger and linking it to the preceding block, thus creating a tamper-proof chain of RUL predictions and model refinements.

(5) Consensus Algorithm: It is the protocol that ensures agreement among nodes in the BC network regarding the validity of transactions and the order in which they are added to the BC. In the context of FL, the consensus algorithm ensures that model updates are accepted or rejected based on agreed-upon criteria. For RUL predictions, the global model only gets updated if most nodes agree that the information, such as the prognostic model's parameters, from a worker is correct and matches what the whole system needs. This communal verification process reinforces the trustworthiness of the RUL predictions.

(6) Monitoring Server: The monitoring server is crucial in overseeing and coordinating the FL process. It provides insights into the performance of participating workers, ensuring that their information contributions are timely and reflect their condition. It also monitors the model's learning progress, ensuring the RUL predictions are as precise as possible. Furthermore, it facilitates the flow of information between the disparate elements of the FL network, ensuring smooth operations and reliable RUL predictions.

These six components together create the structure of the BC-enhanced FL system. This setup allows for safe, confidential, and collective training of models across independent manufacturing sites. The procedure continues, cycling through updates and validations, until the overarching prognostic model has stabilized or achieved the desired level of accuracy. This is crucial for RUL prediction, where precise model convergence ensures reliable forecasts of equipment lifespan, which is essential for maintenance planning and resource allocation. The specific workings of the BC-supported FL algorithm are thoroughly detailed in Algorithm 1, with a comprehensive breakdown available in Appendix A.

## 3.3 Technical details of how BC addresses key challenges within FL in RUL prediction context

As discussed above, combining BC technology with FL holds immense promise in the context of RUL predictions for components or systems. BC ensures data privacy, incentivizes data contributions, and supports scalability. It safeguards sensitive RUL data, encourages data sharing, and adapts to diverse systems. Our proposed BC-FL framework addresses these aspects to enhance the accuracy and reliability of RUL predictions. In detail, they are as follows:

- **Trust and transparency:** BC technology offers an unchangeable ledger for recording and verifying every RUL prediction model update. This transparency builds trust among participants and establishes a clear lineage of contributions. Smart contracts implemented on the BC can enforce predefined rules, guaranteeing fairness and transparency in the FL-based RUL prediction process.

  Let $i$ denote a participating worker (i.e., factory or aerospace company) in the BCbased FL process. The architectural framework for worker $i$'s role in the FL-based RUL prediction training process is visually depicted in Figure 5.

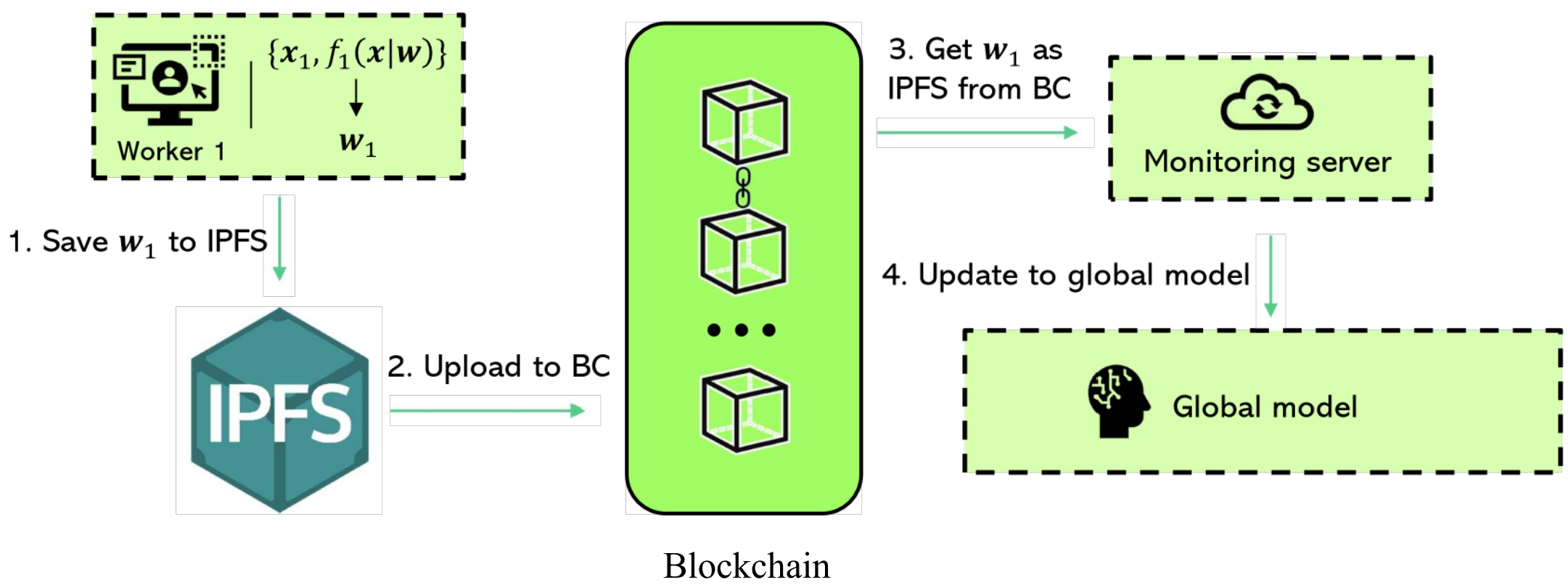

Fig. 5. Architecture for worker $i$'s involvement in the RUL prediction's training process.

This process includes the following steps:

(i) *Save $w_1$ to IPFS*: Assume that worker 1 has successfully trained an ML model for RUL prediction. This worker can then save the model parameters, i.e., $w_1$, to InterPlanetary File System (IPFS) [29]. This step reduces the data form when uploading to BC, i.e., a hash link instead of a big file. In this chapter, we use the NFT. Storage gateway to store the data in decentralized networks.

(ii) *Upload to BC*: Following the successful storage of the model weights $w_1$ on IPFS, in the context of RUL predictions, the next crucial step involves recording the corresponding hash on the blockchain through a SC. This SC is an unalterable ledger, effectively anchoring the model's state to the blockchain. By doing so, we establish an indelible record of the model's weights, introducing transparency and traceability to the FL process, which is especially vital when estimating the RUL of systems. This immutable record ensures that the RUL prediction model's integrity is preserved throughout its lifecycle, enabling reliable and auditable RUL predictions.

The solidity code for uploading the IPFS hash to the BC is summarized in Appendix B.

(iii) *Get $w_1$ as IPFS on BC*: Subsequently, the retrieval of model weights is a critical step. To obtain these weights securely, we invoke a read function within the blockchain smart contract (BC SC). This function provides a secure gateway to access the recorded hash, ensuring the retrieval of the precise model state from the blockchain. This step is pivotal in maintaining the integrity of the FL process, particularly when estimating the systems' RUL. To perform this task, one can harness the web3 library, facilitating interaction with the smart contract and utilizing the read function to fetch the weights as a hash. This process safeguards the RUL prediction model's data and supports the trustworthy derivation of RUL estimates. The Python code snippet to demonstrate this process is shown in Appendix C.

(iv) *Updating the global model*: Following the acquisition of model weights $w_1$, the central server, which oversees the storage of the global model, is ready to incorporate these updated weights. As illustrated in Figure 1, this FL-based RUL prediction process is a pivotal juncture. It amalgamates the collective knowledge derived from various workers (e.g., factories or aircraft companies), a critical aspect when estimating the RUL of machinery or aircraft. This collaborative effort leads to the refinement and heightened accuracy of the global model, ultimately enhancing the precision of RUL predictions.

- **Data privacy and security:** BC technology introduces robust cryptographic techniques, which play a crucial role in bolstering security within the context of systems' RUL predictions. Specifically, these cryptographic methods enable the encryption of data, further fortified by recording access permissions on the blockchain. This combined approach adds a layer of protection to sensitive RUL data. The decentralized security protocol inherent in BC, complemented by its immutable nature, becomes paramount when dealing with RUL predictions. Therefore, it effectively addresses the ongoing challenge of ensuring data privacy and security in federated learning processes dedicated to estimating the systems's RUL.

In integrating BC with FL for data privacy and security, it is imperative to recognize BC's inherent privacy and security features, even within a public BC framework. This characteristic assures that uploaded data, including FL model weights, are inherently protected. However, additional security measures become imperative when applied to the specific requirements of RUL prediction. To fortify data security within RUL prediction, we propose a preemptive measure. We advocate for integrating RSA encryption and decryption techniques, as discussed in Hamza et al. (2020) [30], in conjunction with the hash value before uploading data to the BC. Figure 6 shows that this approach represents data as an IPFS hash link. This comprehensive security strategy ensures that RUL prediction data remains highly safeguarded throughout the FL process, aligning with the stringent privacy and security demands of estimating systems' RUL.

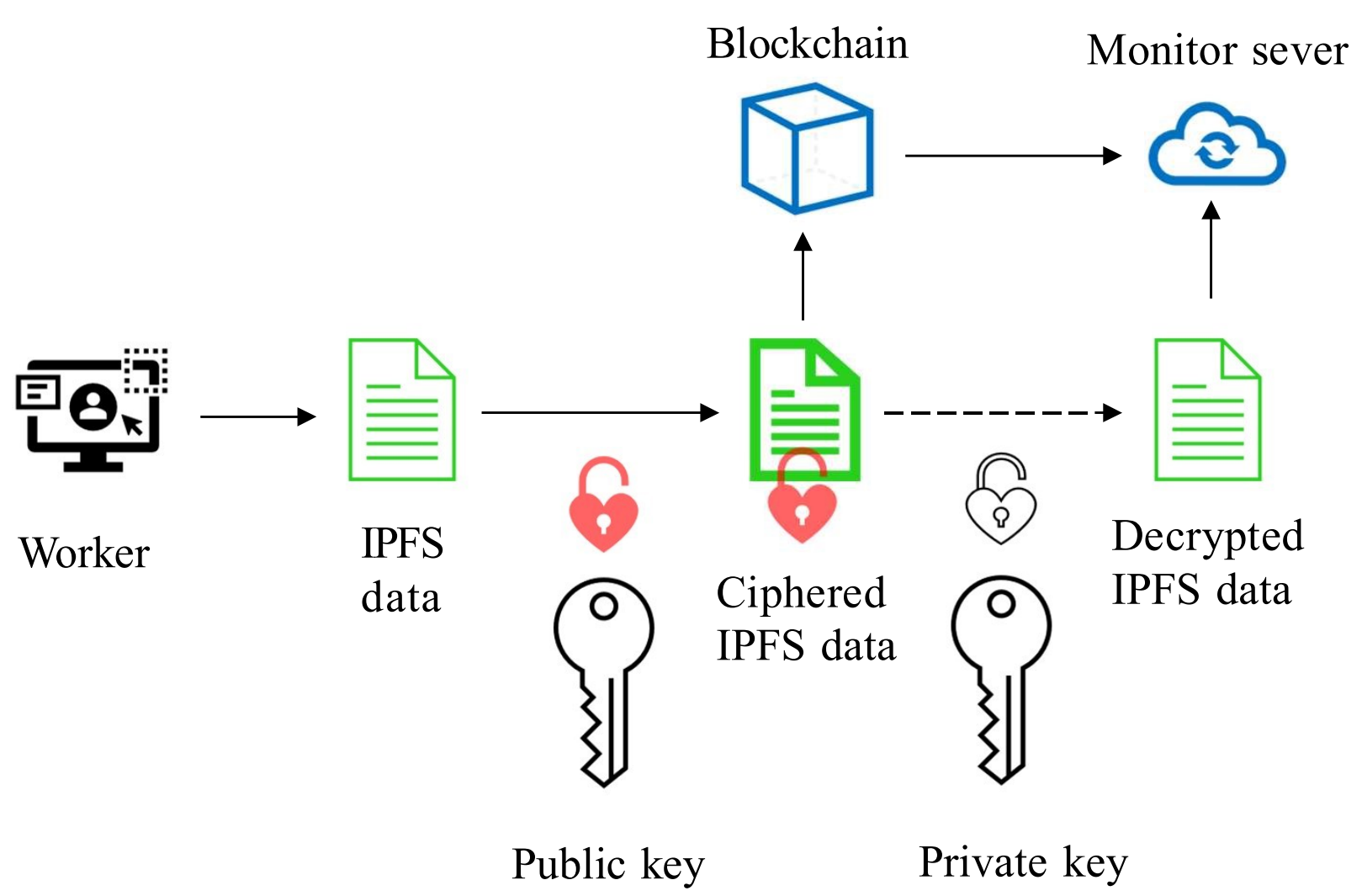

Fig. 6. Integration of RSA encryption and decryption with the hash value before uploading data to the BC as an IPFS hash link.

This approach fortifies data security and establishes an additional layer of protection. It ensures that even if unauthorized parties gain access to the data, they cannot decipher it without the necessary private key. This added security measure is particularly critical when estimating RUL of critical safety systems. Ultimately, this step, illustrated in Figure 6, guarantees that only the central server responsible for global model updates can decrypt the IPFS hash link and retrieve the data. In the context of RUL prediction, this level of data security is paramount, safeguarding the integrity and confidentiality of critical information throughout the FL-based RUL estimation process. The Python code snippet demonstrates the above process, shown in the Appendix D.

- **Incentives and rewards:** BC-based tokens or cryptocurrencies can be integrated to address this issue. SC can automate the distribution of rewards based on the quality and quantity of contributions. Furthermore, a transparent ledger of contributions and rewards on the BC ensures equitable participant compensation. For illustration, when estimating the RUL of machines or aircraft, this approach ensures that factories or aircraft companies are equitably compensated for their valuable input. Beyond monetary incentives, it stimulates active participation and fosters a self-sustaining ecosystem around the federated learning process, further enhancing the accuracy and efficiency of RUL predictions.

The integration of BC brings forth a unique opportunity to implement transparent and immutable incentive structures, addressing a critical aspect often encountered in collaborative ML endeavors. Fig. 7 illustrates the architecture enabling worker $i$ to upload and receive incentives from the BC automatically and seamlessly.

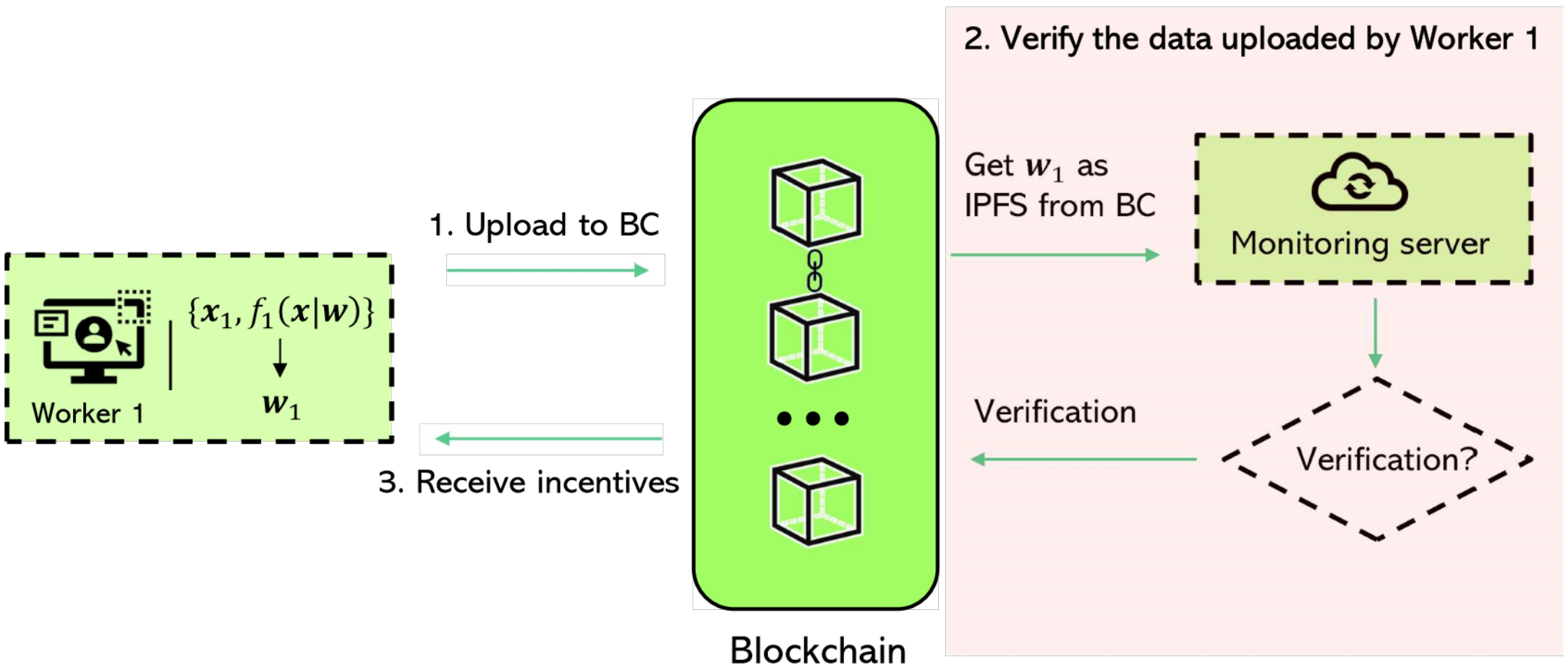

Fig. 7. Architecture enabling worker $i$ to upload and receive incentives from the BC automatically and seamlessly.

This process includes the following steps:

(i) *Upload to BC:* Proceed with the upload of pre-trained weights $w_1$ to the BC, similar to the process outlined in step 1 of Section 3.1. It is imperative to emphasize that, in order to fortify the security and privacy of worker data, we recommend implementing the RSA key, as introduced in Section 3.2, before uploading the IPFS hash to BC.

(ii) *Verify the data uploaded by a worker:* In the context of RUL predictions of critical safety systems, addressing the potential threats posed by malicious data uploads and inadvertent inaccuracies is essential. These actions can potentially distort the global model update, significantly impacting the accuracy of RUL forecasts. While FL's training algorithm offers some protection against such concerns, it may not cover all possible attack vectors that adversaries could exploit. To mitigate the risk associated with erroneous or misleading data uploads, we propose deploying an automated verification service managed by a monitoring server in the backend. This service is crucial in the RUL prediction process, as illustrated in Fig. 7. It involves applying the new weights to the global model and assessing whether they contribute to increased predictive accuracy. If the updated weights enhance the global model's accuracy, it triggers events within the blockchain system, leading to the allocation of incentives to the respective worker. This not only incentivizes genuine contributions but also safeguards against adversarial attempts to manipulate the RUL prediction process, ultimately reinforcing the accuracy and reliability of RUL forecasts in critical safety systems.

(iii) *Receive incentives:* Inspired by the robust cryptographic principles of BC, factories participating in the RUL prediction process can be incethrough tokens [31] or Non-Fungible Tokens (NFTs) [32] on the BC platform. This innovative approach is a powerful motivator, encouraging factories to act by deploying crucial tasks such as verifying and maintaining data integrity. In the context of RUL predictions for critical safety systems, our framework incorporates a monitoring server to assess the quality of worker-contributed prognostic parameters for global model updates. A token or NFT increases the worker through blockchain transactivation. This reward mechanism not only promotes trust but also drives participation and accountability. Subsequently, workers can utilize these tokens or NFTs as coupons or for various relevant rewards offered by the central company overseeing the FL process, further strengthening the commitment to accurate and reliable RUL predictions.

- **Scalability and efficiency** BC technology, with its consensus mechanisms like sharding or sidechains, offers significant enhancements in scalability. The tasks associated with RUL prediction, including computation and validation, can be effectively distributed across the BC network. This distribution alleviates the strain on a centralized server. It allows for the efficient processing of vast amounts of data.Furthermore, BC's integration with SC further streamlines the aggregation process, as illustrated in Figure 8. These SCs automate and optimize data aggregation, significantly enhancing the efficiency of FL. This automation is pivotal for implementing FL in extensive and diverse environments, where RUL predictions play a vital role in optimizing maintenance schedules and resource allocation.

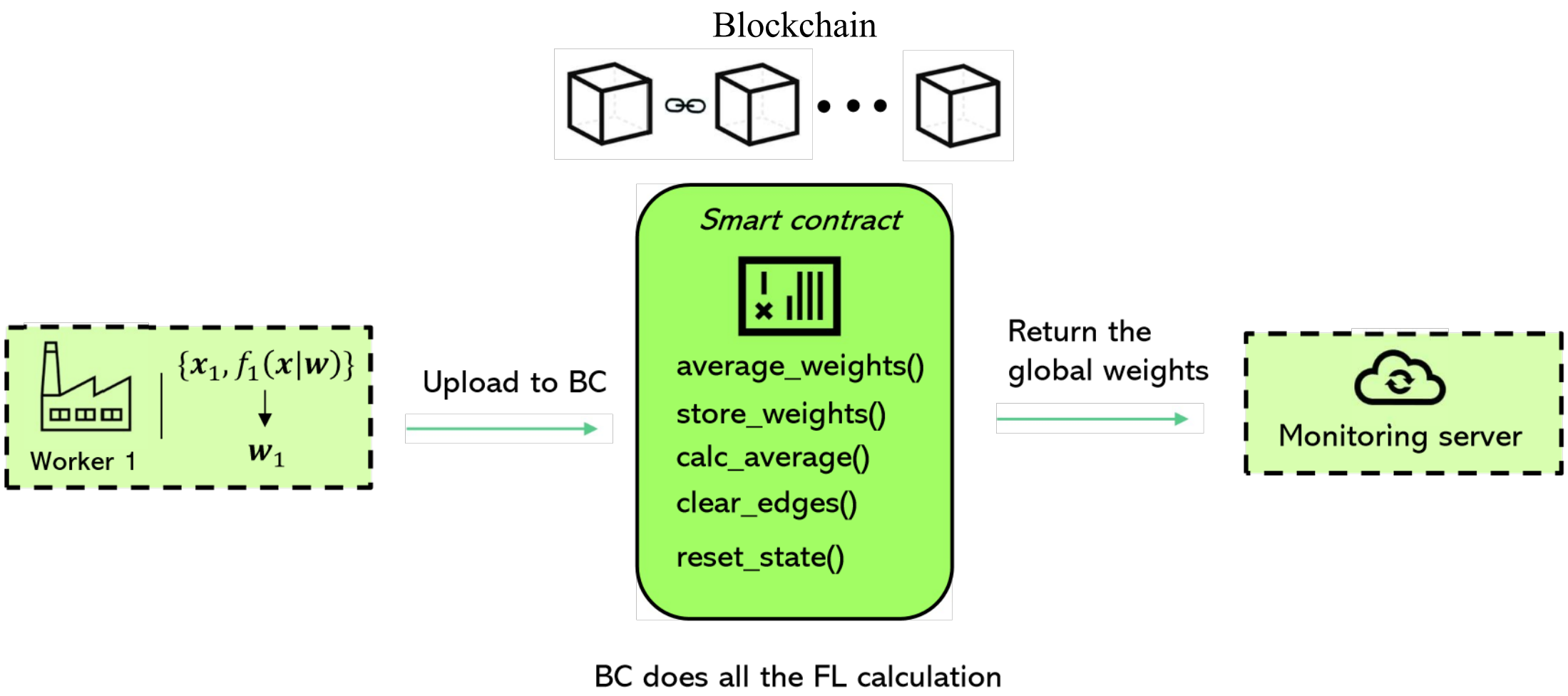

Fig. 8. Proposed BC infrastructure for scalability and efficiency FL-based BC.

Particularly, Fig. 8 shows the proposed BC infrastructure for the scalability and efficiency of FL-based BC. The core idea involves establishing a BC-based infrastructure capable of receiving uploaded weights from participating factories through IPFS hashes and/or an array of numbers. The entire spectrum of FL computations is autonomously executed within this BC network. This encompasses weight averaging, accuracy assessment, and disseminating global weights. Integrating RUL prediction into this process enhances operational efficiency and scalability.

By consolidating these complex calculations within the BC network, as illustrated in Figure 8, we anticipate a significant enhancement in the scalability and operational efficiency of FL endeavors tailored to RUL predictions. This approach streamlines the workflow, optimizes resource utilization, and contributes to more accurate and reliable RUL forecasts in diverse and expansive industrial environments.

# 4 Case study

In this section, we delve into a case study to demonstrate how the proposed framework, a combination of BC and FL, addresses crucial concerns, particularly regarding RUL predictions for critical safety systems. By harnessing the inherent strengths of both technologies, we endeavor to tackle key issues such as trust, data privacy, incentives, and scalability, all of which are pivotal for accurate and reliable RUL forecasts.

Our use case harnesses the synergy between BC's immutable ledger and FL's privacypreserving approach, focusing on their relevance in RUL predictions for aircraft systems. These applications mark a shift towards a transparent, secure, and efficient ML environment tailored for RUL prediction. Our discussion is logically structured for broad accessibility and impact, offering insights into this integrated approach's potential in advancing RUL predictions.

## 4.1 Case study description

This chapter uses the Commercial Modular Aero-Propulsion System Simulation (CMAPSS) dataset developed at NASA Army Research Lab to demonstrate how the proposed framework is efficiently applied to critical safety systems like aircraft engines. This dataset is widely recognized as a benchmark dataset for RUL prediction [23,33].

The C-MAPSS dataset consists of four sub-datasets, i.e., FD001, FD002, FD003, and FD004, each representing different operational conditions and fault patterns. Within each sub-dataset, there are both training and testing datasets. Fig. 9 shows the CMAPSS dataset feature explanation and the illustration of six random sensor data with a full life cycle.

Some column features do not provide useful information for the RUL estimation. As a consequence, 14 of the 21 sensors data and two of the three working conditions, with a total of 16 features, are picked out as the raw input feature to the model as similar to several studies [33–35]. In summary, the picked columns are 3, 4, 6, 7, 8, 11, 12, 13, 15, 16, 17, 18, 19, 21, 24, 25. Hereafter, we introduce the RUL prediction results using the proposed framework.

| Data explanation | |
|---|---|
| Column | Definition |
| 1 | Engine unit |
| 2 | Cycle number |
| 3-5 | Working conditions |
| 6-26 | 21 raw sensor data |

(a)

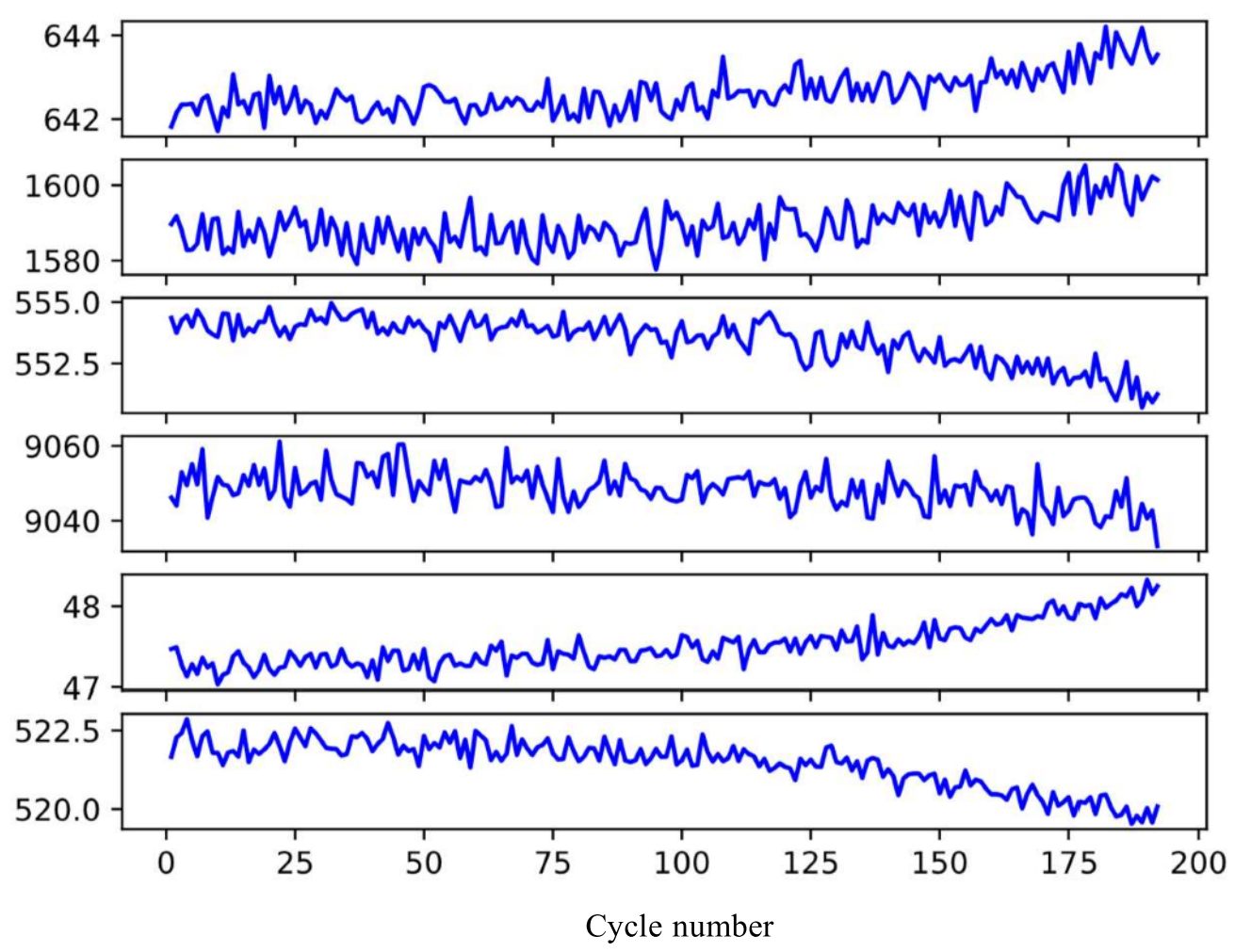

(b)

Fig. 9. (a) CMAPSS dataset feature explanation and (b) Illustration of six random sensor data with a full life cycle.

## 4.2 RUL prediction outcome

This section presents the RUL prediction results using the proposed BC-based FL discussed in Sec. 3.2. This chapter uses the SVM model for the FL process and Binance Smart Chain (BSC testnet) BC to demonstrate the BC-based FL framework. To assess prediction accuracy, we utilize the root mean squared error (RMSE) metric—a widely adopted measure in RUL prediction studies [3,5].

Fig. 10 illustrates the RUL predictions for testing engine units based on the FL-based BC across various time series of CMAPSS. To simplify, we randomly selected one unit each from FD001, FD002, FD003, and FD004 to visualize the results. In general, the synchronous FL-based BC demonstrates a remarkable ability to accurately detect maintenance requirements, even in the case of FD002 and FD004, provided that the machine is halted upon the initial indication of the red line.

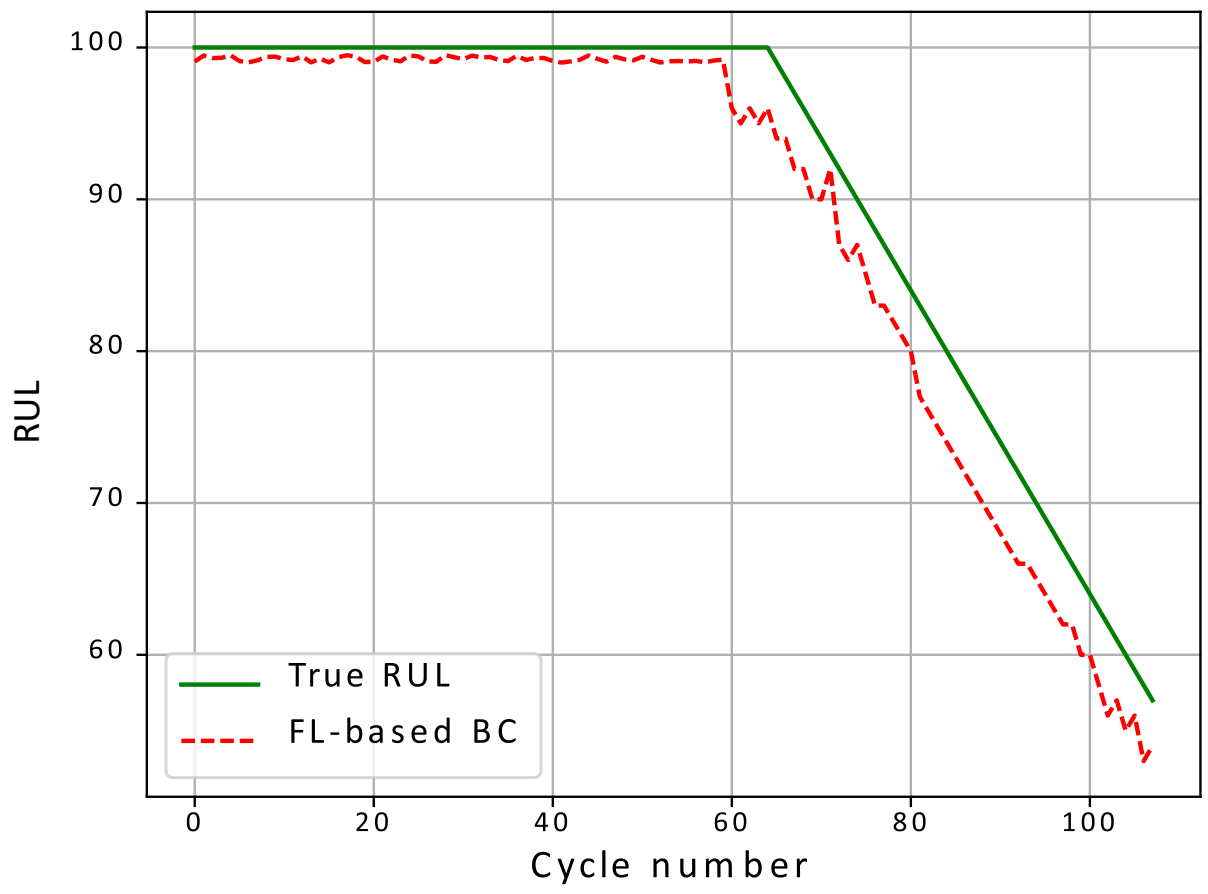

(a)FD 001

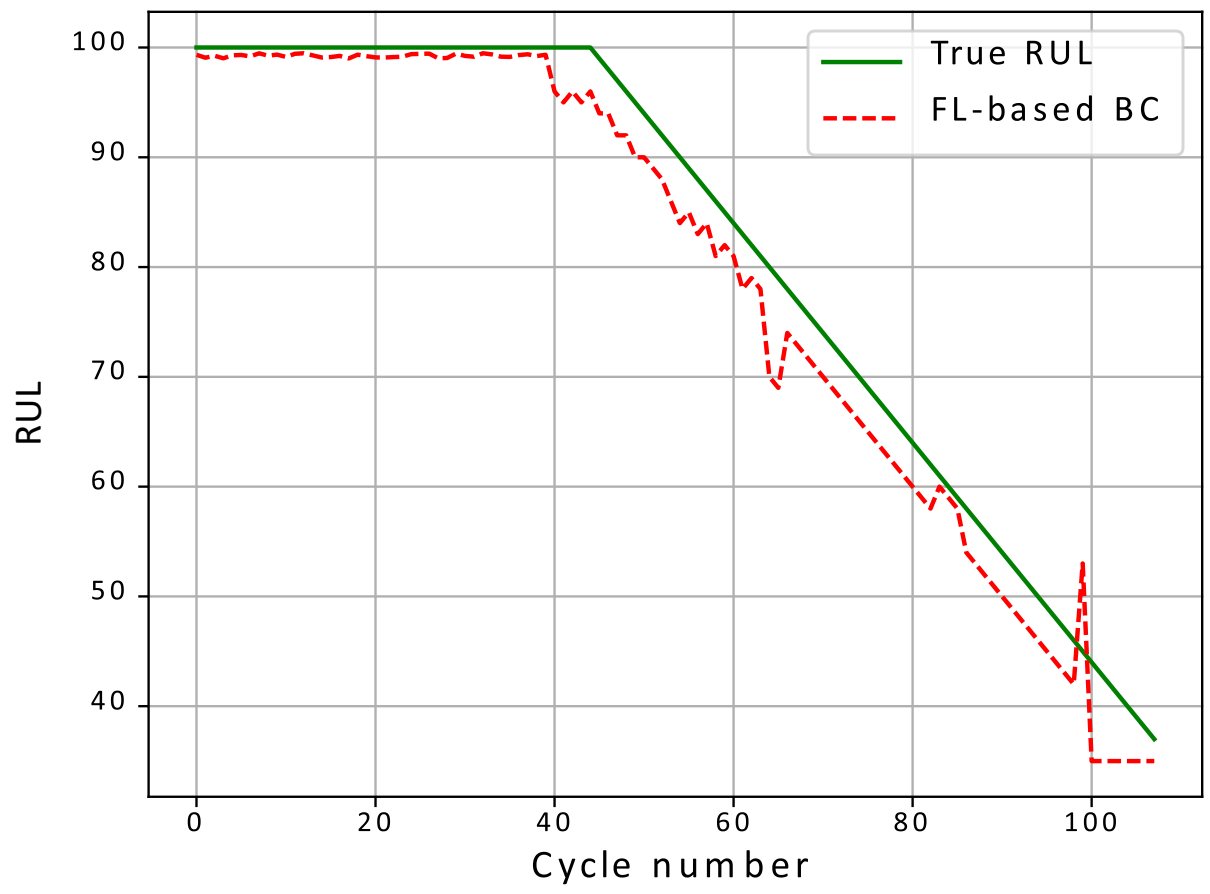

(b)FD 002

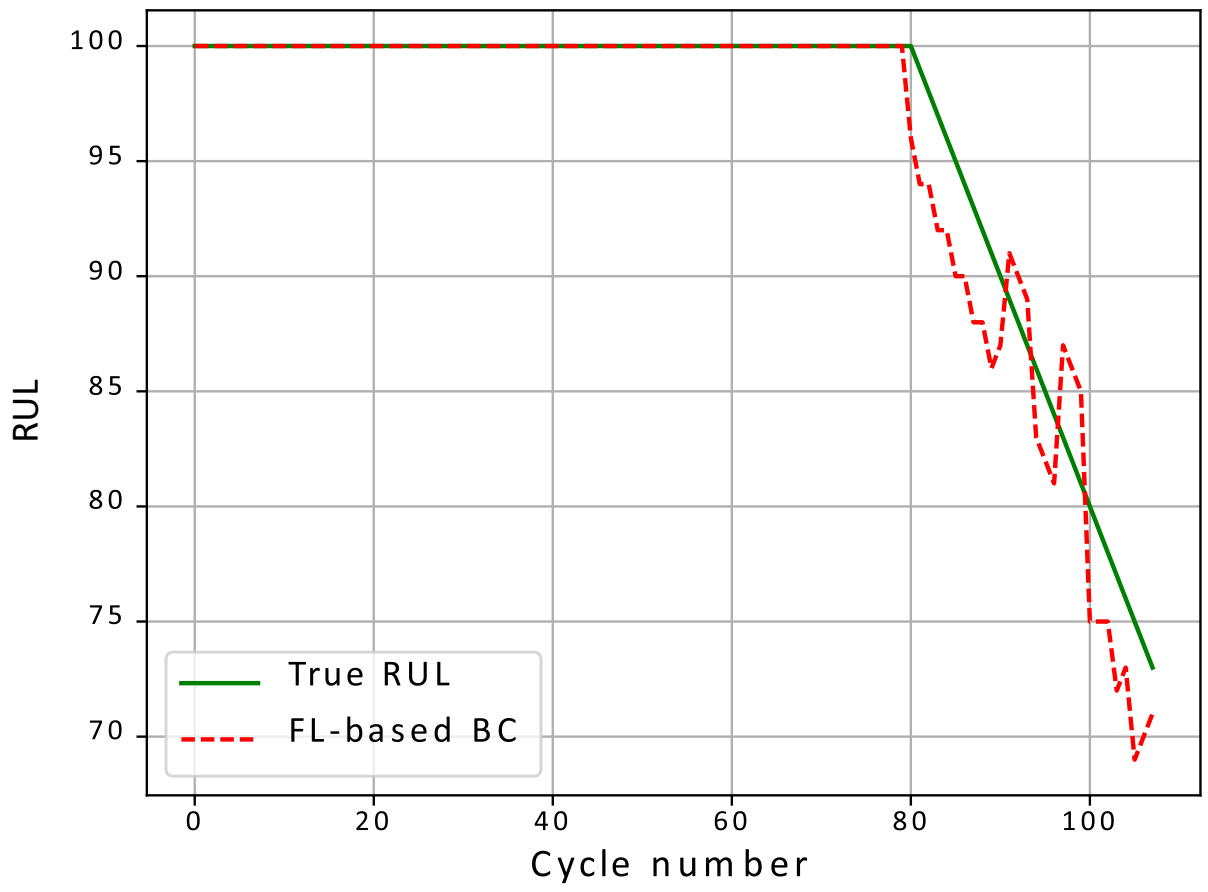

(c) FD003

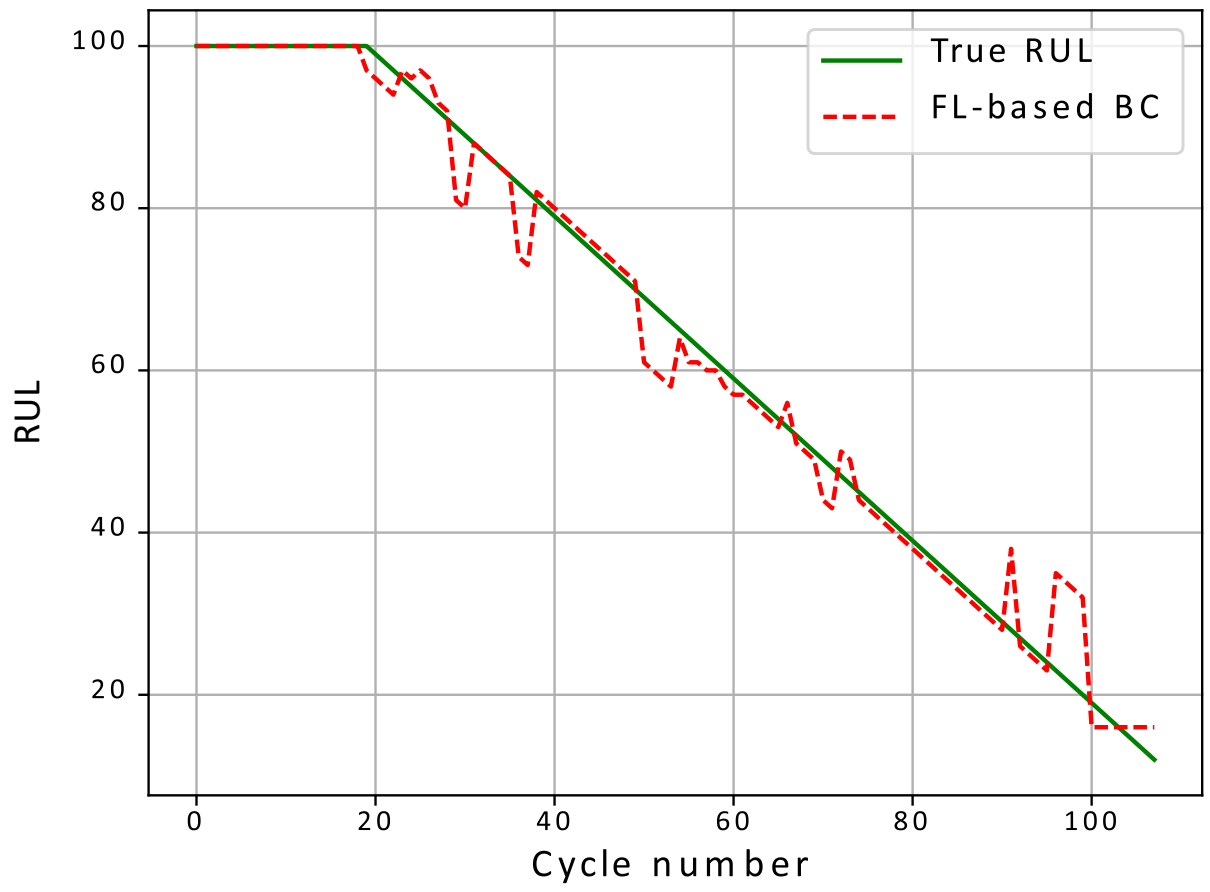

(d) FD004

Fig. 10. RUL predictions for testing engine units based on the FL-based BC across various time series of CMAPSS.

The root mean square errors (RMSE) obtained for each dataset are as follows: 24.76, 32.25, 25.17, and 28.54. In general, the BC-based FL demonstrates a reasonable prediction accuracy, as evidenced by other studies of RMSE [36–38]. The framework addresses the primary challenges associated with FL (refer to Sec. 2.3) and signifies RUL prediction capabilities.

It is noted that the primary scope of this is to introduce a novel method to integrate BC and FL into the domain of RUL prediction to tell the advantages of BC on data security and integrity. We do not aim to improve the accuracy of FL in RUL prediction compared to prior studies using Deep Learning. Integrating FL into BC poses significant challenges, including substantial computational and financial costs. For example, uploading and reading weights from BC can take up to one minute, especially when using complex training models, which could extend to several weeks due to network traffic. Additionally, each weight upload has associated financial costs in gas fees. Consequently, we choose a simpler ML model, such as SVM, to leverage its advantages of simplicity, fast training times, and minimal weight size. This is good enough in the academic world to demonstrate the concept. Therefore, comparing BC-based RUL with those using more complex models as in previous studies [36–38] is not relevant. In perspectives, especially in the industrial world, one should explore advanced fast and light deep learning models that can be combined with BC, leveraging technologies like Oracle or private chains to enhance computational accuracy and reduce financial cost.

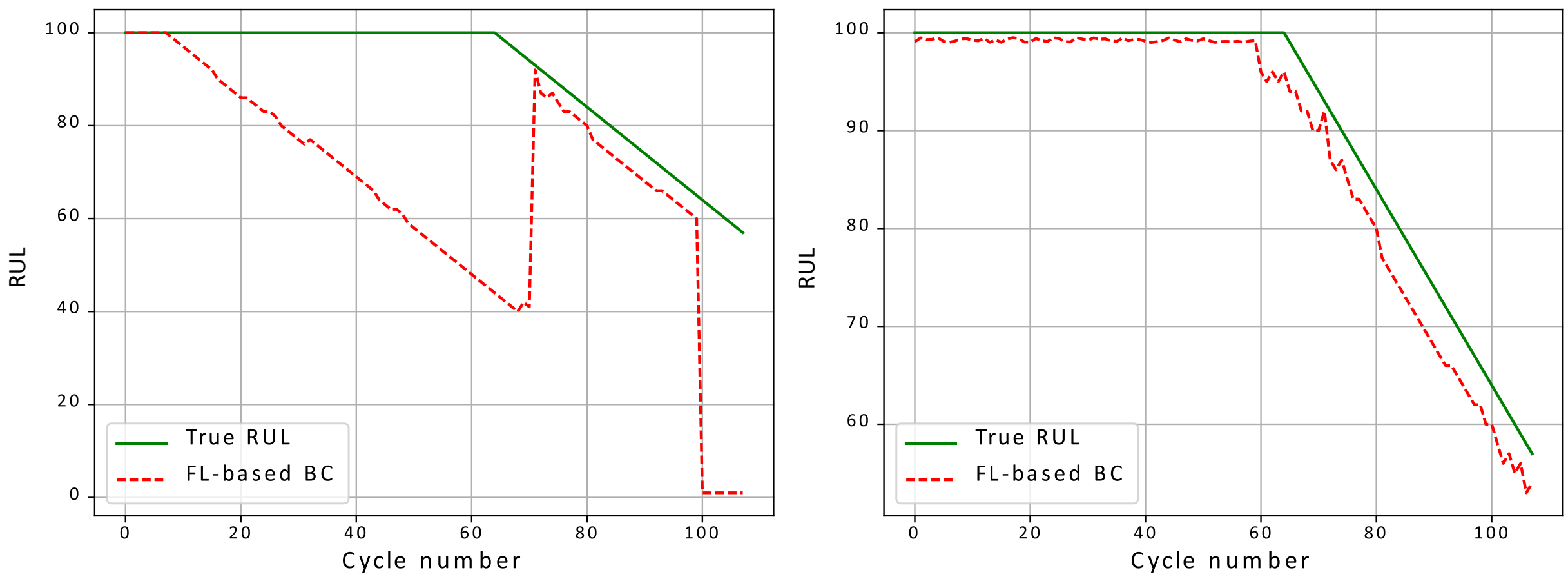

(a) First iteration

(b) Final iteration

Fig. 11. The performance of RUL prediction on the dataset during first and final iteration of FL-based BC.

Using FD001 dataset as an example, Fig. 11 illustrates the performance of RUL prediction on the dataset initially, before FL-based BC. Subsequently, it demonstrates the enhanced RUL prediction accuracy after workers contribute their model weights to the FL process using the BC application. The result indicates a notable improvement in RUL prediction accuracy as workers engage in the FL process, thereby augmenting transparency and privacy due to the inherent nature of BC technology.

## 4.3 Discussion

This section shows the efficiency of our BC-based FL framework in addressing the pivotal challenges outlined in Sec. 2.3 for RUL prediction. We delve into the specific challenges encountered in RUL prediction, introduce a realistic problem, and demonstrate how our BC-based FL framework provides a pertinent solution rooted in the capabilities of BC technology.

### 4.3.1 Trust and transparency

Establishing trust and transparency stands as a critical prerequisite for collaborative model training. The intricacies of this challenge have been thoroughly examined in Sec. 2.3.

**Real world problem statement:** Consider a scenario where multiple aircraft companies aim to collaborate on an FL project to enhance predictive maintenance for their assets. Each company possesses valuable data that, when combined, could significantly improve the accuracy of predictive models for RUL estimation. However, the following challenges may raised from the process:

> • Lack of Trust among Companies: Due to competitive concerns, these companies are reluctant to share their proprietary data directly with each other. There's a lack of trust among the companies regarding safeguarding their sensitive information. • Transparency Concerns: The participating companies may hesitate to fully disclose their internal model updates or operational insights. Concerns arise about revealing too much information about their processes and potentially losing a competitive edge.

**BC solution:** Integrating BC into the FL framework introduces a decentralized and highly secure environment, particularly crucial in the context of RUL predictions for aircrafts. Smart contracts SC play a pivotal role in governing data access and usage, ensuring the protection of sensitive information. This secure approach allows for the collaborative nature of FL while safeguarding confidential data. Furthermore, the BC's transparent and immutable nature significantly enhances trust within the RUL prediction process. It provides an auditable record of model updates and data contributions without divulging proprietary details, as described in Section 3.2. This transparency is visually represented in Figure 14, showcasing the permanence of recorded transactions within the BC. This perpetual record empowers companies to precisely track which data has been disclosed and uploaded by each participating entity, fostering transparency and accountability in the FL process, which is essential for accurate RUL predictions.

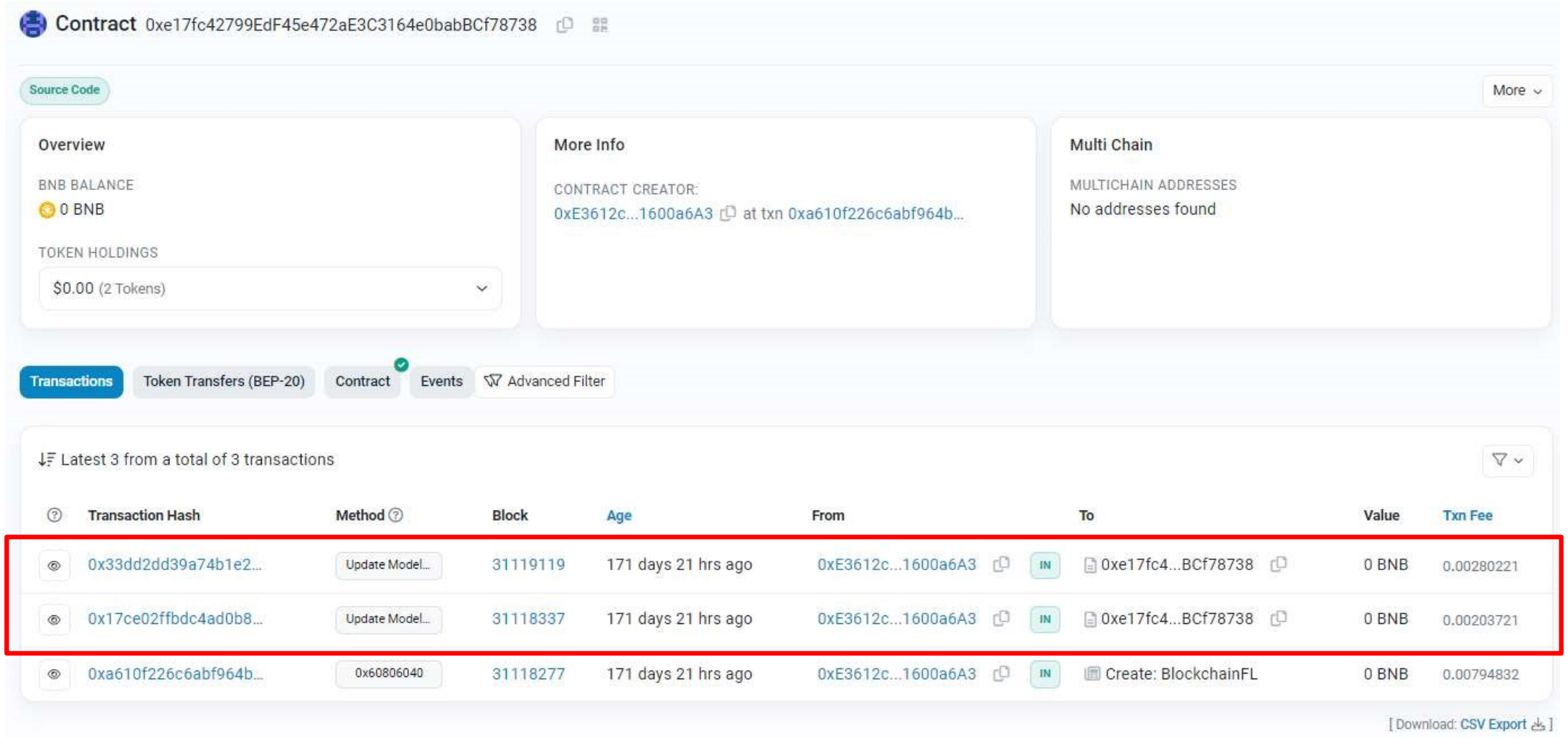

Fig. 12. Recorded transactions that existed forever within the BC. The red rectangle shows that a company has recently uploaded the data to the BC.

In general, this use case highlights the real-world challenges where trust and transparency issues impede the seamless collaboration of companies in an FL network. Implementing a BC-based FL framework can serve as a viable solution to foster trust and transparency while preserving the confidentiality of sensitive information.

### 4.3.2 Data privacy and security

As discussed in Sec. 2.3, ensuring the privacy and security of sensitive information is crucial to adopting FL. The integration of BC with FL presents a promising solution to address these critical concerns by harnessing the immutable and decentralized nature of BC.

**Real world problem statement:** Consider a collaborative effort among aircraft companies from different countries aiming to develop an FL model for predicting an aircraft's RUL. The objective is to leverage diverse datasets to improve the accuracy and generalization of predictive models. However, the following challenges may raised from the process: • Each aircraft manufacturer and operator possesses proprietary maintenance and performance history data. These data are subject to strict privacy and confidentiality regulations. Sharing aircraft-specific information across organizations and borders can raise legal and ethical concerns. This practice can potentially breach data protection laws and industry confidentiality agreements, a matter of significant concern in the aviation sector.

- Security of Sensitive Aircraft Information: Sharing sensitive aircraft data, even in an aggregated form, poses security risks. Aircraft organizations are cautious about potential data breaches during transmission, which could jeopardize the confidentiality of critical information and violate aviation industry regulations.

**BC solution:** The integration of BC introduces advanced cryptographic techniques, notably leveraging the RSA algorithm, to significantly bolster security in the RUL prediction process. Aircraft data undergoes encryption using RSA keys, a well-established cryptographic method renowned for its secure data transmission [30]. Access permissions are meticulously documented on the blockchain, reinforcing data protection with an added layer of security.

This decentralized security approach, coupled with the immutable nature of blockchain (BC), directly addresses the ongoing concern of data privacy in RUL prediction using federated learning (FL). By implementing the RSA algorithm, only the FL operator possesses the decryption key, ensuring the confidentiality and security of sensitive aircraft data throughout the FL process. Figure 13 illustrates the encryption and decryption process using the RSA algorithm. The encrypted hash appears as a distinct string compared to the original, reinforcing its privacy and security. Even if unauthorized parties gain access to the data, they will remain unable to decipher it without the required private key, thus ensuring data protection and privacy in the RUL prediction process.

Fig. 13. Original key encrypted and decrypted using RSA algorithm.

```
Original string: https:// bafybeibbnrtxrzsanqfgifl4awixxvyr53te3jmlwdaxaeifncsnykhgxe.ipfs. nftstorage.link
Encrypted string: b'\xa5\xb5/\xa77\xfb\xb1\xc5e\xaa\xe6\x7f\xb5\x90gj\
    xc8\xb4\xdd\x08+\xd9\xfb\x91\xbe\xc9\xad\xf5w\x90$uw\x91F!\x9f_\xae\
    xb8\x80\x826L\x992p\xe8\xeaJ\x1bxSYoo\xf9)\x1d\xe2\x1d"AM\xf4\xd1\xa1
    \xcc\xf9\x0el\x9b\xde\xdf\xb4ko\xfc\xa8\xdb\x12\x12\x8d \xfagc\xec\
    xe8\x00'H\xb3\x9b>\xda\x0b\xe8\xf5\xbd\xc2Z\x9cB\x914\xcb\x98p\xbej\
    x9f\x90>\x0cu\xc4Q\x07o\x8bo0\x94[\xf8e\xd6'
Decrypted string: https:// bafybeibbnrtxrzsanqfgifl4awixxvyr53te3jmlwdaxaeifncsnykhgxe.ipfs.
    nftstorage.link
```

### 4.3.3 Incentives and rewards

Incentivizing active participation and rewarding contributions are key to building a sustainable ecosystem for RUL prediction through FL. In this section, we explore how BC can be used to create innovative incentive mechanisms, encouraging stakeholders to engage in the FL process actively.

**Real world problem statement:** Imagine an FL process aimed at optimizing predictive maintenance for aircraft. Multiple aircraft operators and maintenance organizations contribute data from diverse sources to enhance the efficiency of RUL predictions. Nevertheless, the context presents the following incentive and reward challenges:

- Establishing Incentives for Data Contributors: Aircraft operators and maintenance organizations investing resources and sharing crucial data may hesitate to participate actively without clear incentives or rewards. The absence of proper motivation can lead to organizations prioritizing their individual interests over collaborative contributions, potentially impacting the accuracy and effectiveness of RUL predictions. • Fair Distribution of Rewards: The challenge lies in devising a fair method for distributing rewards or benefits from improved RUL predictions. Entities involved may worry that their contributions could be undervalued or overlooked during the reward allocation process.

**BC solution:** In predicting aircraft RUL, the BC-based FL framework introduces transparent and automated SC to manage incentive structures. These SCs can programmatically define and execute reward distributions based on each entity's contributions. The decentralized and immutable nature of the blockchain ensures transparency and trust in the reward allocation process. However, the implementation requires an automated verification service in the backend, overseen by a monitoring server (as described in Section 3.2). This service evaluates the impact of new model weights on global accuracy. If the updated weights prove beneficial, events within the blockchain are triggered to allocate incentives to the respective worker, thus promoting collaboration and accuracy in RUL predictions.

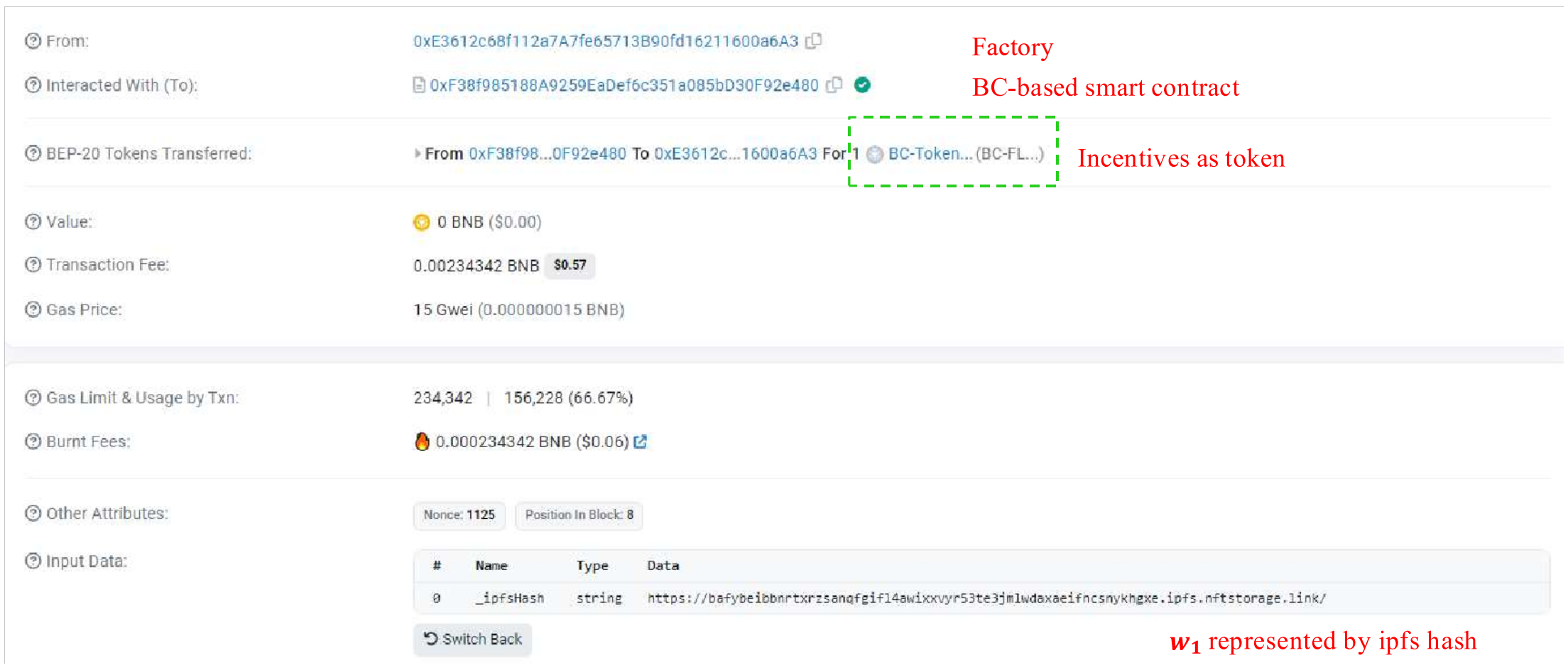

Fig. 14. Incentives transaction details of worker after uploading $w_1$ to BC via SC.

Fig. 14 illustrates the specifics of the incentive transaction for the worker following the upload of $w_1$ to the BC via SC. As depicted in the lower section of the figure, once $w_1$ is uploaded as the IPFS hash, the SC autonomously initiates the transfer of one token, referred to as BC tokens, to the worker's address. Consequently, the worker gains the ability to utilize this token as a coupon or a relevant reward, as facilitated by the central company overseeing the FL process. This incentivizes active participation and contributes to the accuracy of RUL predictions.

#### 4.3.4 Scalability and efficiency

In the rapidly evolving landscape of FL, achieving scalability and operational efficiency emerges as a critical issue. The integration of BC introduces a transformative paradigm that addresses these pivotal concerns. By leveraging the inherent strengths of BC, the FL can seamlessly scale to accommodate diverse and dynamic datasets by optimizing resource allocation and enhancing computational efficiency in BC infrastructure.

**Real world problem statement:** The collaborative endeavor involving multiple aircraft operators and maintenance organizations may give rise to the following scalability and efficiency challenges:

> Scalability in the context of aircraft's RUL prediction poses challenges, particularly when involving diverse aircraft operators and maintenance organizations. As the number of participating entities grows, managing the scalability of the FL system becomes intricate. Coordinating diverse datasets and computational resources becomes more challenging, hindering efficient model updates across a vast and varied company network.In the context of aircraft's predictive maintenance, achieving computational efficiency within the FL process is paramount for real-time decisionmaking. Inefficient model training or communication protocols could result in delays when responding to dynamic shifts in aircraft maintenance needs and operational demands.

**BC solution:** BC-based FL framework introduces decentralized consensus mechanisms and SC to enhance scalability. SC can automate and optimize the coordination of model updates, ensuring efficient collaboration among many participants. The decentralized nature of the blockchain helps distribute the computational load, improving overall efficiency.

It is essential to clarify that, in this chapter, the framework is in its conceptual phase and has not undergone empirical validation. As such, the results of its implementation are not included in this chapter. Instead, we invite the research community and industry experts to provide valuable feedback on this proposed infrastructure. This collaborative effort aims to refine and streamline the integration of BC and FL, ultimately fostering more accessible and impactful applications, particularly in the aircraft industry.

In general, the four challenges discussed in Sec. 4.3 demonstrate the envisioned BCenabled FL infrastructure, representing a promising solution for revolutionizing collaborative FL paradigms. By unifying the strengths of BC and FL, we anticipate a transformative impact on industries that seeking secure, efficient, and scalable solutions.

# 5 Conclusion

This chapter introduces an innovative paradigm that combines FL and BC technology to tackle the challenges of predictive maintenance in diverse global industry settings. By safeguarding data privacy, ensuring trust, and addressing operational nuances, this approach offers an efficient solution for RUL prediction. The main contributions of the chapter are: • This first study integrates BC technology into FL for RUL prediction, adopting an accessible approach tailored to a diverse audience across various industrial domains. Additionally, we have made our open-source software and code available on GitHub, encouraging collaborative contributions from the scientific community to develop further and advance this promising framework (https://github.com/tqdpham96/blockchainfederate-learning-use-cases).

- The proposed framework is showcased within the manufacturing field, specifically focusing on RUL prediction of aerospace engineering. The outcomes of RUL prediction using the CMAPSS dataset effectively tackle the key challenges linked with FL and underscore the framework's robust capabilities in predicting RUL compared to other studies.

- An efficiency assessment of the proposed BC-based FL framework is performed, searching into specific challenges encountered in RUL prediction for multiple industrial fields. Addressing a realistic problem scenario, we demonstrate how our BC-based FL framework offers a pertinent and effective solution, leveraging the capabilities inherent in BC technology.

In the context of Industry 5.0, where collaboration between humans, AI, and machines takes center stage, this approach can drive significant improvements in operational efficiency, resilience, and sustainability. By securely integrating AI-driven prognostics into decentralized networks, we aim to enable predictive maintenance solutions that are more accurate and aligned with Industry 5.0's emphasis on human-centric, adaptive, and intelligent manufacturing processes.

Moreover, future work could enhance transparency and interpretability within the BCbased FL framework. Explainable AI (XAI) is critical in providing insights into how AI models make decisions, which is especially important for industries requiring high levels of trust and accountability, such as manufacturing and aerospace. By incorporating XAI techniques, we can offer clear justifications for RUL predictions and anomaly detection results, making it easier for stakeholders to understand model behavior and trust the system's recommendations. Additionally, refining anomaly detection methods using advanced AI techniques could help identify rare and unforeseen patterns within decentralized manufacturing environments. This would enable early identification of potential issues, reducing downtime and improving the resilience of Industry 5.0 systems. Combining XAI with anomaly detection will ensure that the decision-making process is both robust and transparent, aligning with the evolving needs of smart and human-centric manufacturing ecosystems.

## 6 Acknowledgements

This chapter was supported by the grant (IAD-4) funded by the International Research Institute for Artificial Intelligence and Data Science (IAD), Dong A University, Vietnam

# Appendices

## A BC-based FL's algorithm

**Algorithm 1** BC-based FL's algorithm

$K$: number of factory
$f_i(w)$: the factory $k$'s loss
$M_j$: BC miner
**procedure** BCFL($M_j$, $K$)
    **for** each factory $i$ in $K$ **do**
        $w_i \leftarrow$ set of pre-trained weight's ML model of factory $i$
        upload $w_i$ to BC by any valid types
        Miner $M_j$ start mining process by consensus algorithm
        Adding a new block to the current BC
        Download the global weight $w$
        **return** $w$ to factory $i$
    **end for**
**end procedure**

---

## B Solidity-based smart contract for uploading the weights as the IPFS hash to the BC

```
contract BlockchainFL { mapping(address => mapping(uint => string )) public
    ipfsMappingAddress; // get ipfs hash from SC
    mapping(address => uint) public addressUploadedTime;

    function updateModelHash( string memory
        _ipfsHash
    ) external { require( msg.sender != address(0),
                "Invalid address"
        );
        uint timeUploaded = addressUploadedTime[msg.sender];
        ipfsMappingAddress[msg.sender][timeUploaded + 1] = _ipfsHash;
        addressUploadedTime[msg.sender] += 1;
    }
}
```

# C Python code to read web3 transaction

```
# Import the necessary libraries from web3 import
Web3

# Please replace with your node’s URL, contract address, API, and factory address to test the code.

# Connect to the Ethereum node
w3 = Web3(Web3.HTTPProvider(’http://localhost:8545’))
# Define the contract address and ABI contract_address =
’0x1234567890abcdef1234567890abcdef12345678’ contract_abi = [
    # Define your contract’s ABI here
]
# Create a contract instance contract = w3.eth.contract(address=contract_address, abi=contract_abi)
# Define the function to retrieve weights from SC def
get_weights_from_contract(address):
    weights_hash = contract.functions.ipfsMappingAddress(address).call() return weights_hash # Example
usage weights_hash = get_weights_from_contract() print(f"Weights Hash: {weights_hash}")
```

## D Python code to perform rsa algorithm

```
# Import the necessary libraries import rsa

# Get public key and private key public_key, private_key =
rsa.newkeys(1024)
# Define the ipfs hash link
```

```
message = "https://ipfsabcdefgh123.ipfs.nftstorage.link"
# Encrypt and decrypto the ipfs hash enc_message = rsa.encrypt(message.encode(),
public_key) dec_message = rsa.decrypt(enc_message, private_key).decode()
# Example usage
print(f"Original IPFS Hash: {message}") print(f"Encrypted IPFS Hash: {enc_message}")
print(f"Decrypted IPFS Hash: {dec_message}")
```